\documentclass[a4paper,man,floatsintext,longtable,noextraspace,12pt]{apa6}
\usepackage[english]{babel}
\usepackage[utf8x]{inputenc}
\usepackage{amsmath}
\usepackage{graphicx}
\usepackage[colorinlistoftodos]{todonotes}
\usepackage{hyperref}

\usepackage{booktabs}
\usepackage{longtable}
\usepackage{array}
\usepackage{multirow}
\usepackage{wrapfig}
\usepackage{float}
\usepackage{colortbl}
\usepackage{pdflscape}
\usepackage{tabu}
\usepackage{threeparttable}
\usepackage{threeparttablex}
\usepackage[normalem]{ulem}
\usepackage{makecell}
\usepackage{xcolor}
\newlength{\cslhangindent}
\newenvironment{cslreferences}%
  {\setlength{\parindent}{0pt}%
  \everypar{\setlength{\hangindent}{\cslhangindent}}\ignorespaces}%
  {\par}

\title{\textbf{Transparency to hybrid open access through publisher-provided metadata: An article-level study of Elsevier}}
\shorttitle{Hybrid OA Elsevier}
\threeauthors{Najko Jahn}{Lisa Matthias}{Mikael Laakso}
\threeaffiliations{Göttingen State and University Library, University of Göttingen\\
Platz der Göttinger Sieben 1, 37073 Göttingen, Germany\\
najko.jahn@sub.uni-goettingen.de
}{Department of Political Science, John F. Kennedy Institute, Freie Universität Berlin \\
Lansstraße 7-9, 14195 Berlin, Germany \\
l.a.matthia@gmail.com
}{Information Systems Science, Hanken School of Economics \\
Arkadiankatu 22, 00100 Helsinki, Finland \\
mikael.laakso@hanken.fi
}

\authornote{Correspondence concerning this article should be addressed to Najko Jahn}

\abstract{With the growth of open access (OA), the financial flows in scholarly journal
publishing have become increasingly complex, but comprehensive data and
transparency into these flows are still lacking. The opaqueness is especially
concerning for hybrid OA, where subscription-based journals publish individual
articles as OA if an optional fee is paid. This study addresses the lack of
transparency by leveraging Elsevier article metadata and provides the first publisher-level study of hybrid OA uptake and invoicing. Our
results show that Elsevier's hybrid OA uptake has grown steadily but slowly from
2015-2019, doubling the number of hybrid OA articles published per year and
increasing the share of OA articles in Elsevier's hybrid journals from 2.6\% to
3.7\% of all articles. Further, we find that most hybrid OA articles were
invoiced directly to authors, followed by articles invoiced through agreements
with research funders, institutions, or consortia, with only a few funding
bodies driving hybrid OA uptake. As such, our findings point to the role of
publishing agreements and OA policies in hybrid OA publishing. Our results
further demonstrate the value of publisher-provided metadata improve the
transparency in scholarly publishing by linking invoicing data to bibliometrics.}

\begin{document}
\maketitle

\hypertarget{introduction}{%
\section{Introduction}\label{introduction}}

The rise of open access (OA) has added complexity to
scholarly publishing, particularly concerning transparency of economic
dimensions. Financial transparency in journal publishing has long been
lacking because Big Deal subscription contracts between academic
institutions and large academic publishers usually include
confidentiality clauses that prohibit publicly sharing
the scope or pricing of such agreements (Bergstrom et al., 2014;
Frazier, 2001; Larivière et al., 2015). Further, the distributed payment processes
to access and publish scholarly articles add to the opacity of
the financial flows of scholarly publishing (Lawson et al., 2016) as they can
involve several actors simultaneously, including authors, research
institutions, libraries, and funders.

The problem with this lack of transparency becomes more apparent in the
case of hybrid OA. In this OA business model, individual articles can be
made openly available by paying an article processing charge
(APC), while the journal as a whole remains subscription-access. Hence,
hybrid OA was originally introduced as a low-risk transitional model
that allows journals to gradually convert to full OA and reduce
subscription costs as the OA uptake increases (Prosser, 2003). Many
large subscription publishers have since incorporated hybrid OA, but there have been concerns that the hybrid model allows for double dipping---receiving two payments for one article, the APC and
subscription fees (Prosser, 2015; Shieber, 2009). While publishers
assure they adjust their pricing
(Mittermaier, 2015), without transparency around the uptake of hybrid OA
and both revenue streams, such claims are impossible to evaluate. This is especially concerning because hybrid OA has recently gained popularity after a decade of only slow uptake (Björk, 2012). The main drivers of this development have been research institutions and funders that implemented  OA policies, allocated OA funds, and established
formal agreements with publishers that allow
affiliated authors to publish OA free-of-charge (Laakso \& Björk, 2016).

With the recent introduction of transformative agreements, an evolving
concept describing contracts that shift library spending from subscriptions to OA
(Borrego et al., 2020; Hinchliffe, 2019), the demand for
publisher-provided data has increased. More
transparency about hybrid OA uptake and funding could  facilitate the assessment and adjustment of publisher contracts (Schimmer et al., 2015), enhance OA mandate compliance monitoring, and avoid double-dipping (Larivière \& Sugimoto, 2018). However,  previous studies have noted an absence of transparency in hybrid OA publishing and a considerable lack of
publicly available and standardized data (Laakso \& Björk, 2016; Lawson, 2015; Pinfield et al.,
2016)

This paper investigates to what extent publisher-provided
data can increase the transparency around hybrid OA by taking Elsevier,
the largest scholarly journal publisher 
(Larivière et al., 2015), as the starting point for an empirical
analysis of hybrid OA uptake and invoicing. Our approach is based on
openly available data sources and combines scholarly metadata
from Crossref, a DOI registration agency, with
 APC invoicing data from Elsevier. The
compiled data enable us to analyze the 
hybrid OA uptake in Elsevier journals between  2015 and 2019 by license and subject, and to compare it with
Elsevier's full and delayed OA program. Further, we determine whether
hybrid APCs were waived or charged to authors or as part of publishing agreements. In the latter case, we 
examine the invoiced academic consortia and research funders. As such, our findings have implications for research and OA
policy and highlight the potential of publisher-provided data for
large-scale, instantaneous analysis of hybrid OA uptake and invoicing.

\hypertarget{background}{%
\section{Background}\label{background}}

\hypertarget{prevalence-uptake-studies}{%
\subsection{Prevalence and uptake
studies}\label{prevalence-uptake-studies}}

In the first article-level study on hybrid OA, Laakso \& Björk (2016)
reported that between 2007 and 2013, the five largest
publishers--Elsevier, Sage, Springer, Taylor \& Francis, and
Wiley--recorded growing numbers of hybrid journals that coincided with a
twenty-fold increase in hybrid OA (from 666 to 13,994). As a
result, hybrid journals with at least one OA article more than doubled
in number (1,082 in 2009, 2,714 in 2013) but fell by 31\% relative to
the total number of hybrid journals. Laakso and Björk's
(2016) exploratory approach yielded a comprehensive overview tracking the growth of hybrid
OA over time and across multiple publishers, but due to the amount of
manual data collection and cleaning it
is unsuitable for repeated use. Minimizing such manual tasks,
Nelson \& Eggett (2017) narrowed their focus to one publisher--the
American Chemical Society (ACS)--and requested a list of all hybrid OA
articles published from 2006-2011 from the ACS. To assess the
hybrid OA uptake, the authors aggregated and compared the number of OA
(n=814) and subscription articles (n=27,621), finding that over the five
years, 2.9\% of ACS articles were published as hybrid OA.

Contrary to the previous two studies, Kirkman (2018) took a research
funder as a starting point, reporting a 26\% share of hybrid OA for
articles funded by the Australian National Health and Medical Research
Council from 2012-2014 (816 of 3,190). Kirkman collected research
articles using the Funding Agency and Funding Text search fields on Web
of Science, and classified articles as OA when they were freely
available on the publisher website, and further distinguished hybrid OA
articles based on journal-level information from the publisher website,
such as hybrid OA journal and APC lists. Another approach is presented
by Pölönen et al. (2020), who used current research information system
(CRIS) data---institutional publication data---to study the extent of OA
among Finnish research publications from 2016-2017 (n=34,507 research
articles). As part of the analysis, the authors identified hybrid OA
articles through a dedicated OA status metadata field that has been
mandatory to report in Finland since 2016 and found a 7\% share of
hybrid OA. However, the drawbacks of Kirkman's (2018) and Pölönen et
al.'s (2020) approaches are that such data sources might not be openly
available or, regarding CRIS data, might not be available at all because
not all countries maintain comprehensive institutional publication data.

Other studies assessed the share of (hybrid) OA for the entire corpus of research articles. Using DOI-assigned research articles indexed in the Web of Science Core Collection
(n=2,610,305) as a benchmark, Martín-Martín et al. (2018) investigated how
many research articles from 2009 and 2014 are freely available through
Google Scholar. After identifying OA articles through licensing
information from Crossref, the authors estimated the share of hybrid OA
at around 0.5\% in 2009 and 1.5\% in 2014. Similarly, in a study assessing the OA level among all scholarly articles and that experienced by Unpaywall users, Piwowar et
al. (2018) obtained license information through Crossref and web-scraping. The study found 3.6\% of 100,000 random Crossref DOIs were hybrid OA
and pointed towards a growing trend in recent years that recorded a
hybrid OA share of 9.4\% of all articles published in 2015.

Since its launch, Unpaywall has become widely used in bibliometric research and rankings (Huang et al., 2020; Robinson-Garcia et al., 2020). However, the lack of standardized and comprehensive publisher-provided data has required several updates to Unpaywall to improve hybrid OA identification and differentiation from  delayed OA (Piwowar et al., 2019; Unpaywall, n.d.-b; Unpaywall, n.d.-c), illustrating ongoing challenges in tracking and comparing hybrid OA prevalence over time.
%Unpaywall recently launched the paid service Unsub that allows libraries to evaluate journal subscriptions by combining
% OA uptake rates with internal financial and usage data (Chawla, 2020).

\hypertarget{financial-studies}{%
\subsection{Financial studies}\label{financial-studies}}

Pinfield et al.'s (2016) analysis of APC payment records provided by 23
United Kingdom (UK) higher education institutions revealed a sharp increase in central
payments from 2007-2013, which was largely attributed to the
introduction of block grants by Research Councils UK (RCUK) and
non-compliance sanctions by the Wellcome Trust. Moreover, the study
showed that OA fees were paid almost exclusively through block grants (92\%), and only a small number of APCs were paid
through internal funding (7\%). In contrast, a recent Springer Nature
survey found that authors draw on a range of funding sources to cover OA
fees, such as dedicated institutional OA funds, block grants, OA
agreements, or research grants (Monaghan et al., 2020). Most hybrid OA
authors were supported through dedicated institutional OA funds (43\%,
excluding block grants) and OA agreements with Springer Nature (34\%).
The differences between these two studies might be due to different policy and funding arrangements---considering the introduction of OA
agreements since Pinfield et al. (2016) and Monaghan et al.'s
(2020) more regionally diverse sample.

Regional and policy differences in APC payments also came to light in a
study by Jahn \& Tullney (2016) that analyzed APC records from 30 German
higher education and research institutions, the Austrian Science Fund (FWF), Jisc, and the Wellcome
Trust. In particular, the study revealed large differences in the amount
of hybrid OA funded from 2014-2015. Whilst hybrid OA accounted for less
than 1\% of APCs paid by German institutions (23 of 3,846), the three
non-German research funders recorded a hybrid share of 75\% (11,533 of
15,779). According to Jahn \& Tullney (2016), this could point towards
differences in science policy, such as hybrid OA being supported by the
three non-German research funders but not by Germany's largest national
funder, the German Research Foundation (DFG). Another possibility is that German hybrid OA fees were
paid from budgets not reported to the Open APC initiative, a crowd-sourcing effort (Pieper \& Broschinski, 2018) from where
the authors acquired data. Among these unreported funds
are research grants and research unit budgets, which author surveys
identified as APC funding sources (Graaf, 2017; Monaghan
et al., 2020). As such, Jahn and Tullney's (2016) findings could reflect
the complexities and potential limitations of institutional OA spending
data that Pinfield et al.~(2016) and Monaghan et al.~(2020) attributed
to incomplete or missing records.

In recent years, national consortia in Europe negotiated publishing agreements covering hybrid OA fees for affiliated authors (Borrego et al., 2020). While these improved invoicing workflows, internal assessments of hybrid OA uptake and invoicing remain challenging because transparent and comparative data have largely remained absent (Marques \& Stone 2020). Such publicly available information might be scarce because publishers lack or withhold such data, or due to confidentiality clauses (Marques \& Stone 2020; Marques et al., 2019; Monaghan et al., 2020).

\subsection{Research questions and aims}

In this paper, we focus on Elsevier, a prominent example in recent hybrid OA uptake and financial studies (Laakso \& Björk 2016; Pinfield et al. 2016). Elsevier's OA portfolio presents the challenges in examining hybrid OA described above. For instance, distinguishing between different OA types. Elsevier supports delayed (Elsevier, n.d.-b), hybrid, and full OA, including so-called mirror journals---full OA counterparts of hybrid journals addressing OA policies opposed to hybrid OA (Harrison, 2019). Further, Elsevier processes APC invoices through various channels, such as agreements with research funders and library consortia or, in the absence thereof, the authors (Elsevier,
n.d.-a, n.d.-c). Surprisingly, we found Elsevier article-level metadata embedded in XML full-texts indicating the articles' OA status and invoicing.  Leveraging this publicly available data, we 
address the lack of transparency around hybrid OA noted in previous studies.
In particular, we use this novel approach to answer the following
questions:

\begin{itemize}
\item
  What was the uptake of Elsevier's hybrid OA publishing option between
  2015 and 2019?
\item
  Through which channels were hybrid APCs invoiced, and who were the
  recipients?
\end{itemize}

\hypertarget{methodology}{%
\section{Methodology}\label{methodology}}

\begin{figure}[H]

{\centering \includegraphics[width=0.95\textwidth]{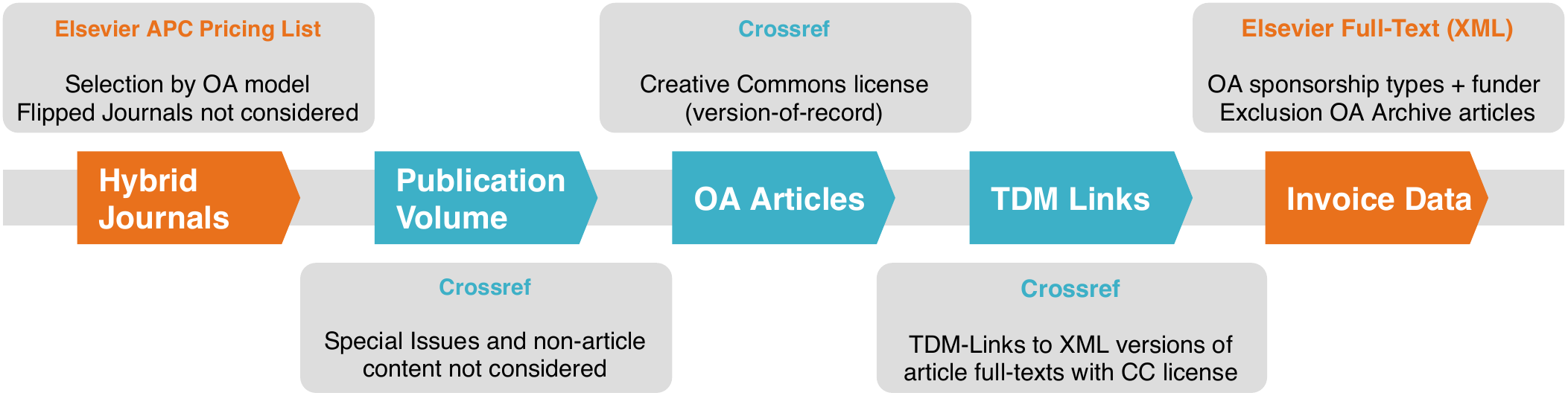}

}

\caption{Data Collection Workflow to Obtain Article-Level OA and Invoicing Data.
}\label{fig:workflow}
\end{figure}

For this study, we collected data relating to Elsevier's hybrid OA
option by drawing on multiple freely available data sources. We
identified Elsevier hybrid journals through Elsevier's APC list and
supplemented our sample with Crossref metadata and text-mined invoicing
data to investigate the invoicing of immediate OA articles provided
under a Creative Commons (CC) license in subscription-based journals. Figure
\ref{fig:workflow} visualizes the automated workflow we used to collect
data from Elsevier and Crossref.

First, we identified hybrid journals through an Elsevier APC price list
from May 2020 provided by Matthias (2020). We excluded journals that transitioned from hybrid to full OA and
reverse-flip journals that flipped 
from full to hybrid OA (Matthias et al., 2019). Overall, we identified 1,970 unique hybrid journals that published between 2015 and 2019.

Next, we used an openly available Crossref database snapshot (Crossref,
2020), which contains all Crossref records registered until March 2020,
to calculate the journals' combined article volume for the five-year
period from 2015-2019. We only included articles published in regular
issues aside from supplements containing conference contributions like
meeting abstracts, indicated by non-numeric pagination. We excluded
non-scholarly journal content, such as the table of contents, following Unpaywall's paratext recognition approach
(Unpaywall, n.d.-a), which we expanded to include patterns indicating
corrections. Furthermore, we categorized the articles by subject
according to the All Science Journal Classification code (ASJC) and
added citation indicators from the Scopus Source Title list (version
June 2020) based on the journal they were published in.

We identified OA articles
through CC licenses in Crossref metadata records (Hendricks et al., 2020) and then downloaded the XML
version of all CC-licensed articles published in a hybrid journal. From
the XML files, we obtained the articles' OA status and invoicing information (see Table \ref{tab:els_xml}). We determined whether articles were immediate or delayed
OA using the XML node \texttt{openArchiveArticle} and measured the
uptake of hybrid OA. Moreover, we obtained the invoice channels and recipients of hybrid OA APCs. Using the \texttt{openaccessSponsorType} node, we distinguished between four invoice channels, including
invoices billed to authors, as part of publishing
agreements with funding bodies (cf.~Elsevier
(n.d.-a)), exempted through fee waivers (e.g., in ``cases of genuine need'' or due to
society or university sponsorships, cf.~Elsevier,
n.d.-c), and other types not
specified by Elsevier. Finally, when hybrid APCs were invoiced as part
of agreements, we identified invoice recipients through the
\texttt{openaccessSponsorName} node.

\begin{table}[H]
\caption{\label{tab:els_xml}Metadata in Elsevier XML Full-Texts.}
\centering
\begin{tabular}[t]{lp{10cm}}
\toprule
XML node & Description\\
\midrule
\texttt{openaccessArticle} & Was the article open
access? \\
\texttt{openaccessSponsorType} & Invoice channel.\\
\texttt{openaccessSponsorName} & Invoice recipient.\\
\texttt{openArchiveArticle} & Was open access provided through
Elsevier's Open Archive program?\tabularnewline \\
\bottomrule
\end{tabular}
\end{table}

We manually classified invoice recipients based on their institutional
sectors, countries, and primary research areas. Following the OECD's
Frascati Manual (OECD, 2015, p. 91), we coded for four sectors: business
enterprise, government, higher education, and private non-profit. 
Due to the low article volume,
we combined the business enterprise sector with invoice recipients Elsevier listed as ``authors'' and
``third-party sponsor''
 into ``Others''. Moreover, we categorized invoice
recipients according to the countries representing their
scope of funding and based on the following primary research areas
health sciences, life sciences, physical sciences and mathematics,
social sciences and humanities, broad (i.e., multiple
research areas), and unknown. Further, we compared Elsevier's invoicing data
with institutional spending data from the Open APC initiative.

Throughout this mostly automated data gathering and analysis process, we
used tools from the Tidyverse (Wickham et al., 2019) for the R
programming language (R Core Team, 2020). To allow for efficient data
manipulation and retrieval, we imported the Crossref dump to Google
BigQuery, applying the rcrossref (Chamberlain et al., 2020) parsers to
extract relevant metadata fields. We used crminer
(Chamberlain, 2020) to obtain the XML-full texts from Elsevier.

\hypertarget{results}{%
\section*{Results}\label{results}}
\addcontentsline{toc}{section}{Results}

This section first presents the results of our analysis of hybrid OA
uptake in Elsevier's journal portfolio with a view to licensing,
disciplinary differences, and citation impact. Then, we present a
descriptive analysis of Elsevier's invoicing data, highlighting
licensing and disciplinary differences for invoicing channels and invoice recipients.

\hypertarget{uptake-of-hybrid-open-access}{%
\subsection*{Uptake of Hybrid Open
Access}\label{uptake-of-hybrid-open-access}}
\addcontentsline{toc}{subsection}{Uptake of Hybrid Open Access}

\hypertarget{overview}{%
\subsubsection*{Overview}\label{overview}}
\addcontentsline{toc}{subsubsection}{Overview}

Between 2015 and 2019, 1,755 of 1,970 (89.1\%) hybrid journals published
at least one OA article. Together these journals published 71,643 OA
articles, which represented 3\% of their total article volume
(n=2,422,087). Table \ref{tab:overview} presents findings for each
year and the aggregated five-year period, illustrating moderate growth
of hybrid OA over time. Each year, the number of hybrid journals
with at least one OA article increased, as did the number of immediate
OA articles in these journals, which nearly doubled from 10,672 in 2015
to 19,311 in 2019. However, since the journals' total article output also grew over time, the relative share of OA articles
in hybrid journals only increased slightly from 2.6\% to 3.7\%.

\begin{table}[H]

\caption{\label{tab:overview}Elsevier Hybrid OA Uptake  2015-2019.}
\centering
\begin{tabular}[t]{lrrrrrr}
\toprule
  & 2015 & 2016 & 2017 & 2018 & 2019 & 2015-19\\
\midrule
\addlinespace[0.3em]
\multicolumn{7}{l}{\textbf{Elsevier Hybrid Journals with ≥ 1 OA article}}\\
\hspace{1em}Total & 1,317 & 1,364 & 1,401 & 1,501 & 1,600 & 1,755\\
\addlinespace[0.3em]
\multicolumn{7}{l}{\textbf{Articles in Elsevier Hybrid Journals with ≥ 1 OA Article}}\\
\hspace{1em}Total & 406,701 & 422,423 & 436,418 & 468,952 & 518,062 & 2,422,087\\
\hspace{1em}Avg. & 308.8 & 309.7 & 311.5 & 312.4 & 323.8 & 1,380.1\\
\hspace{1em}SD & 401.1 & 419.6 & 427.5 & 445.6 & 479.9 & 1,937.5\\
\addlinespace[0.3em]
\multicolumn{7}{l}{\textbf{OA Articles in Elsevier Hybrid Journals with ≥ 1 OA Article}}\\
\hspace{1em}Total & 10,672 & 12,729 & 13,361 & 15,570 & 19,311 & 71,643\\
\hspace{1em}Avg. & 8.1 & 9.3 & 9.5 & 10.4 & 12.1 & 40.8\\
\hspace{1em}SD & 12.9 & 19.7 & 15.6 & 16.2 & 20.1 & 68.9\\
\addlinespace[0.3em]
\multicolumn{7}{l}{\textbf{OA Uptake in Elsevier Hybrid Journals with ≥ 1 OA Article}}\\
\hspace{1em}Total (\%) & 2.6 & 3.0 & 3.1 & 3.3 & 3.7 & 3.0\\
\hspace{1em}Avg. & 3.7 & 4.3 & 4.4 & 4.7 & 5.4 & 3.8\\
\hspace{1em}SD & 4.5 & 5.9 & 5.1 & 4.9 & 5.7 & 4.1\\
\bottomrule
\end{tabular}
\end{table}

\begin{figure}[H]
\caption{Open Access Uptake per Elsevier Hybrid Journal (Per Year).}\label{fig:boxuptake}
{\centering \includegraphics[width=0.7\linewidth,]{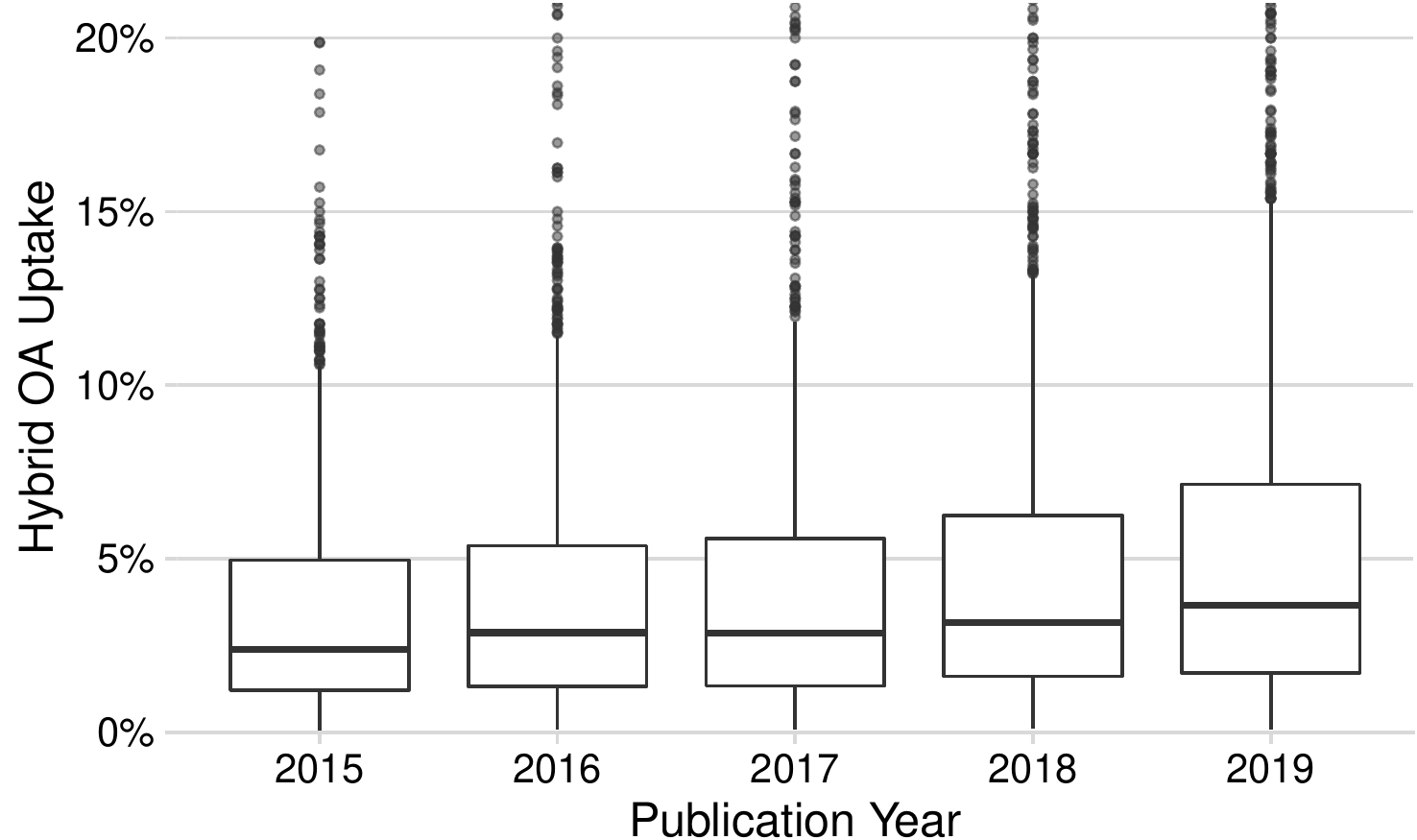} 

}
\vspace{1ex}

     {\raggedright \textit{Note:} The Y-axis is limited to an OA share of 20\%. \par}

\end{figure}

Figure \ref{fig:boxuptake} presents a box-and-whiskers plot of the OA uptake in Elsevier hybrid journals by year,
showing variation among the journals. Notably, the upper quartiles and
whiskers stretch farther over the years, which indicates that the range
of uptake among journals with an above-average proportion of OA articles
increased over the years. Nevertheless, the prevalence of hybrid OA
remained relatively low; in 2019, 95\% of Elsevier hybrid journals
published ≤15.6\% of their articles OA.

\begin{figure}[H]

\caption{Elsevier OA Article Volume 2015-2019 by (A) OA Type and (B) Open Content License.}\label{fig:oa_volume_fig}

{\centering \includegraphics[width=0.7\linewidth,]{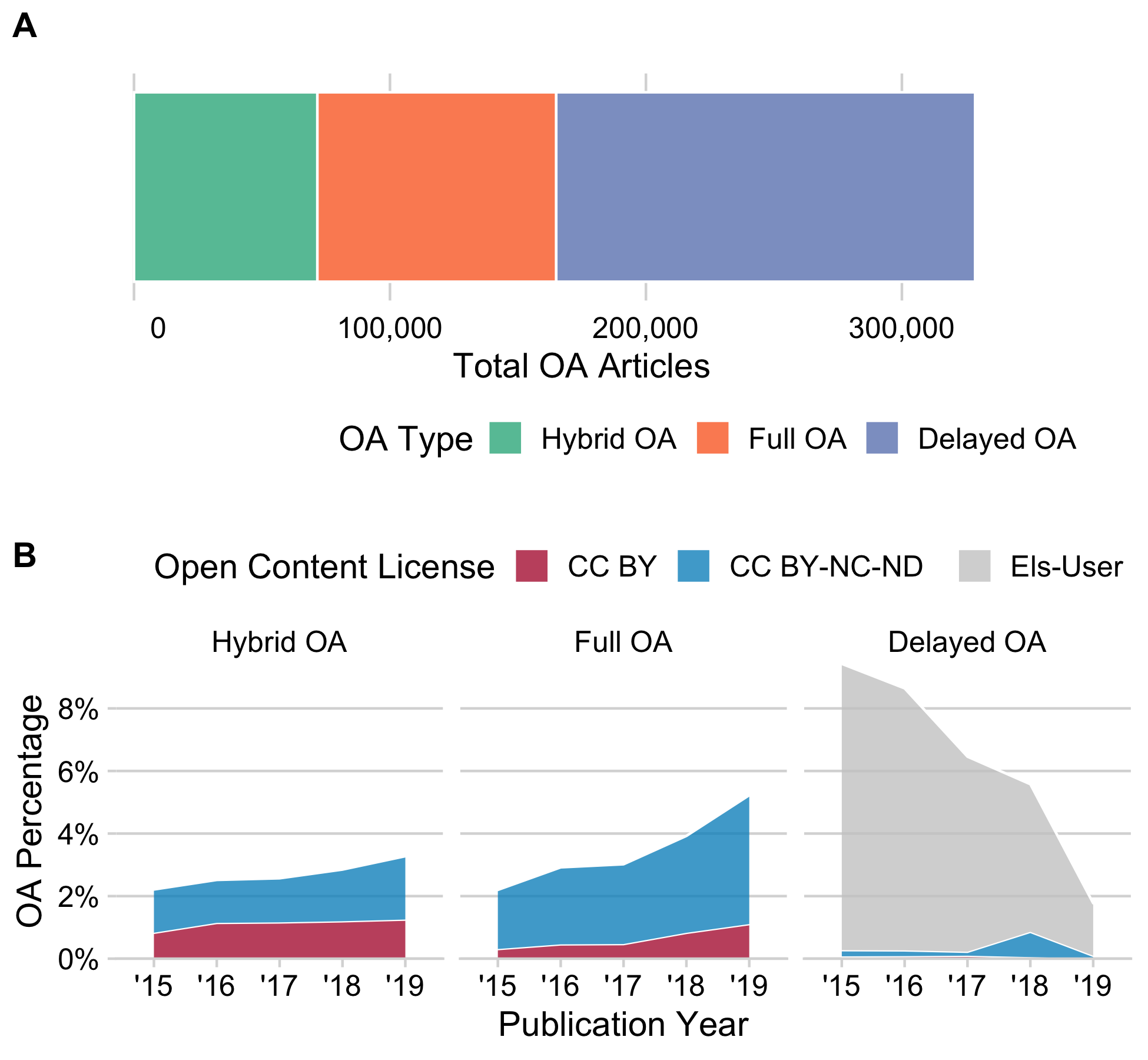} 

}

\end{figure}

During the same period, Elsevier provided OA to 328,601 articles in total
(i.e., full, hybrid, and delayed OA combined), representing 12.4\% of the total
article volume (n=2,643,474). Looking at Figure \ref{fig:oa_volume_fig},
it is apparent that the OA article volume of hybrid journals (n=71,643;
21.8\%) lagged behind delayed OA
(n=163,643; 49.8\%) and full OA journals (n=93,315; 28.4\%), which
included 817 articles published in 38 mirror journals.

\hypertarget{license-prevalence}{%
\subsubsection*{License Prevalence}\label{license-prevalence}}
\addcontentsline{toc}{subsubsection}{License Prevalence}

As far as we observed, OA articles in Elsevier hybrid journals were
published under two possible CC licenses--CC BY or CC BY-NC-ND. CC BY
allows others to distribute, remix, adapt, and build upon the licensed
work, including for commercial purposes, as long as the original author
is credited. In contrast, CC BY-NC-ND is less permissive as it prohibits
commercial reuse and derivatives. From
2015-2019, the proportion of hybrid OA articles published under a CC
BY-NC-ND license marginally increased and maintained a greater share
than CC BY (see Figure \ref{fig:oa_volume_fig}-B).
Interestingly, Figure \ref{fig:oa_volume_fig}-B also reveals that nevertheless hybrid
journals had the highest number and proportion of OA
articles licensed under CC BY (n=29,752; 41.5\%) compared to full OA
journals (n=17,293; 18.5\%) and Elsevier's delayed OA program (n=1,568;
1\%). Most delayed OA articles were provided under an Elsevier user
license (Els-User), which prohibits reuse for commercial purposes.

\hypertarget{subject-area-and-field}{%
\subsubsection*{Subject Area and Field}\label{subject-area-and-field}}
\addcontentsline{toc}{subsubsection}{Subject Area and Field}

Next, we present the hybrid OA uptake of different disciplines. Table
\ref{tab:subject_area_table} presents the high-level findings by ASJC
subject area. Between 2015 and 2019, the physical sciences
(712 hybrid journals; 29,584 OA articles) and health sciences (634
hybrid journals; 25,119 OA articles) recorded the most
hybrid journals with at least one OA article. However,
the life sciences (538 hybrid journals; 31,383 OA articles) published the most hybrid OA articles, while the
social sciences published the least (372 hybrid journals; 11,204 articles). It is important to
note, though, that there is a large overlap between the journals' subject areas.
Therefore, Table \ref{tab:subject_area_table} also assigns hybrid
journals and OA articles fractionally, where journals assigned more than one ASJC subject were counted once for each. For the
remaining results, we only present the full counting.

\begin{table}

\caption{\label{tab:subject_area_table}Full and Fractional Counting of Subject Areas (2015-2019).}
\centering
\begin{tabular}[t]{lrrrr}
\toprule
\multicolumn{1}{c}{ } & \multicolumn{2}{c}{Full counting} & \multicolumn{2}{c}{Fractional counting} \\
\cmidrule(l{3pt}r{3pt}){2-3} \cmidrule(l{3pt}r{3pt}){4-5}
Subject area & Journals & OA articles & Journals & OA articles\\
\midrule
Health sciences & 634 & 25,119 & 506.6 & 18,415\\
Life sciences & 538 & 31,383 & 368.5 & 21,753\\
Physical sciences & 712 & 29,584 & 584.4 & 23,915\\
Social sciences & 372 & 11,204 & 283.5 & 7,462\\
\bottomrule
\end{tabular}

\vspace{1ex}

   {\raggedright \textit{Note:} One multidisciplinary journal (61 OA articles) and 11 journals (37 OA articles) that  could not be matched to the Scopus Source Title list to obtain ASJC codes are not included.\par}

\end{table}

\begin{figure}[H]

\caption{OA Uptake in Elsevier Hybrid Journals by Subject Area and Field (2015-2019).}\label{fig:oa_sub_uptake}

{\centering \includegraphics[width=0.9\linewidth,]{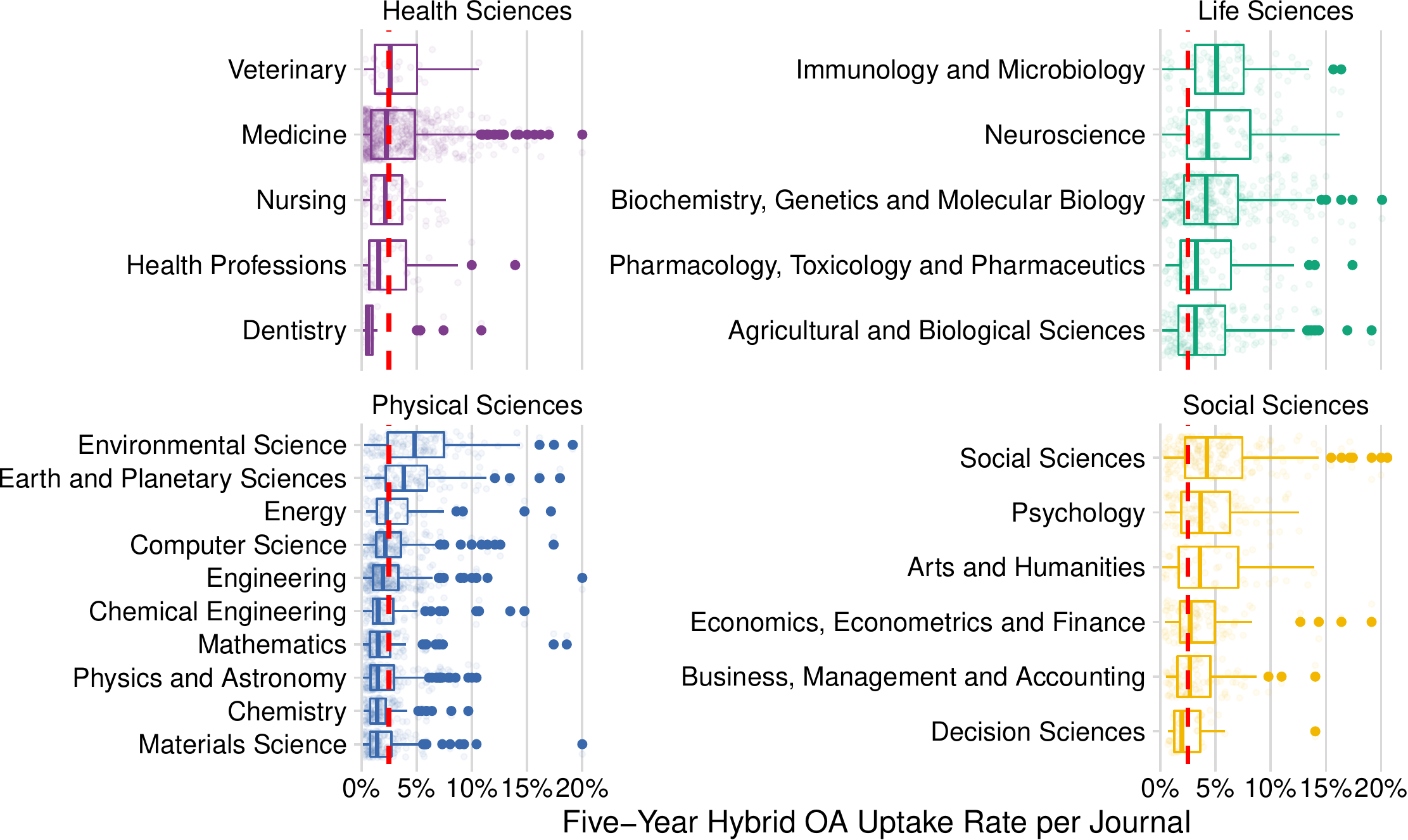} 

\vspace{1ex}

   {\raggedright \textit{Note:} Y-axis is limited to an OA share of 20\%; the red line represents the overall median (2.5), jittered dots represent individual journals. Subject fields are ordered based on the median OA uptake.\par}
   
}

\end{figure}

Figure \ref{fig:oa_sub_uptake} presents the journals' five-year OA
uptake grouped by subject area and field and in comparison to the
overall median (Mdn= 2.5\%; excluding one journal coded as
multidisciplinary). The box plot reveals some variation between the
subject areas. Notably, with the exception of environmental science and
earth and planetary sciences, hybrid journals from the physical sciences show a
lower median OA proportion, whereas most life
sciences and social sciences journals ranked above average. Figure X
also shows large variations among the 26 subject fields, with the
highest OA rates in immunology and microbiology (Mdn=5.1\%), environmental
science (Mdn=4.8\%), and neuroscience (Mdn=4.3\%), whilst chemistry
(Mdn=1.4\%), materials science (Mdn=1.4\%), and dentistry (Mdn=0.6\%) recorded
the lowest OA rate.

\hypertarget{citation-impact}{%
\subsubsection*{Citation Impact}\label{citation-impact}}
\addcontentsline{toc}{subsubsection}{Citation Impact}

\begin{figure}[H]

\caption{SNIP versus OA Uptake in Elsevier Hybrid Journals 2019.}\label{fig:sniptest}

{\centering \includegraphics[width=0.7\linewidth,]{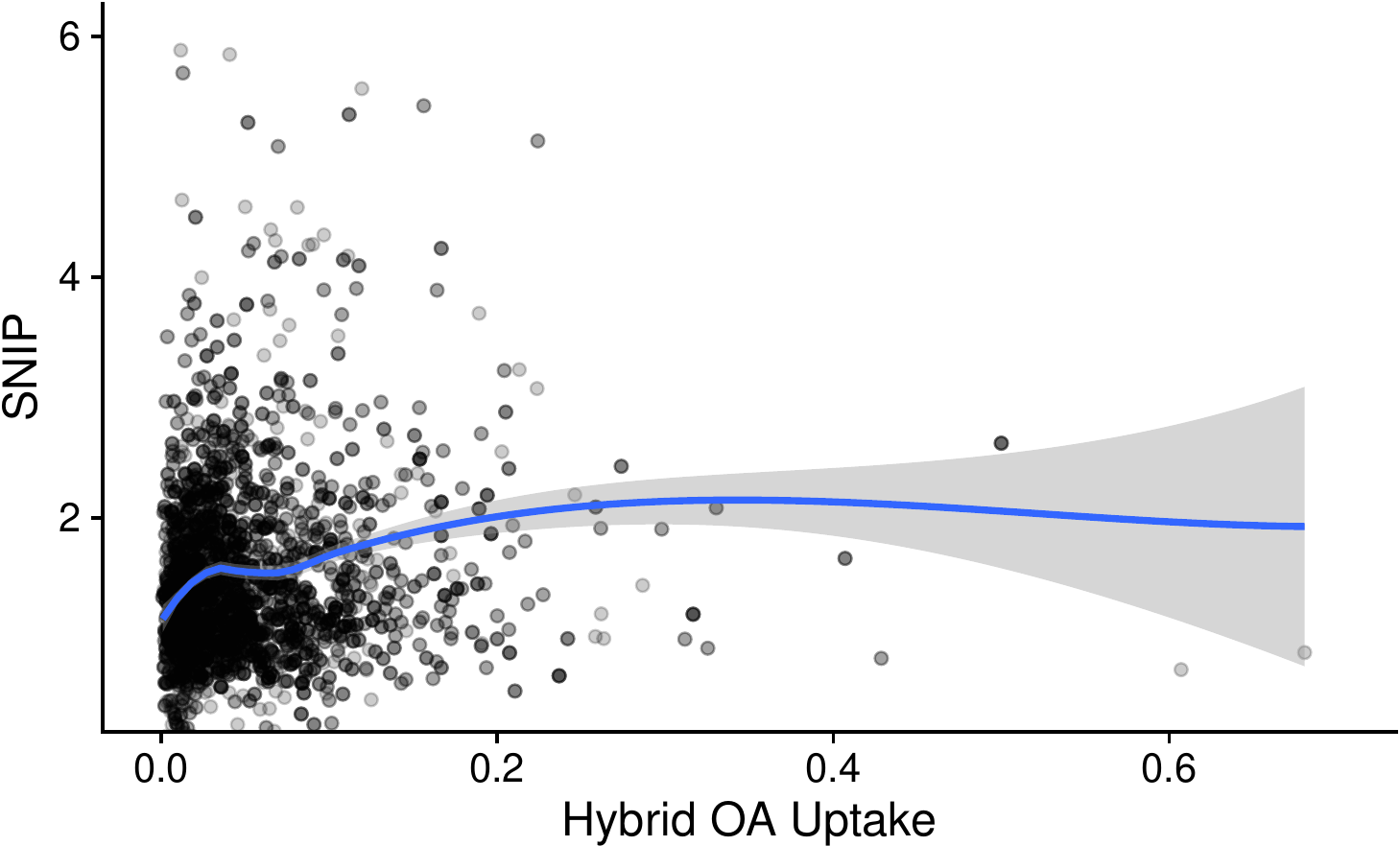} 

}
\end{figure}

Furthermore, we compared the
journals' field-specific citation impact and their OA uptake. Spearman's rho
correlation coefficient was used to assess the relationship between the
2019 Source Normalized Impact per Paper (SNIP) value calculated by
Scopus and the journals' OA uptake that year. Considering only journals
with at least one OA article, we found a weak but positive correlation
between journal impact and OA uptake (\(r_s\)=0.1679, \(p<0.001\)),
suggesting that the relationship is not very strong, based on our data
(see Figure \ref{fig:sniptest}).

\hypertarget{invoicing-for-hybrid-open-access}{%
\subsection*{Invoicing for Hybrid Open
Access}\label{invoicing-for-hybrid-open-access}}
\addcontentsline{toc}{subsection}{Invoicing for Hybrid Open Access}

\hypertarget{invoice-channels}{%
\subsubsection*{Invoice Channels}\label{invoice-channels}}
\addcontentsline{toc}{subsubsection}{Invoice Channels}

As can be seen from Table \ref{tab:invoicing_overview}, hybrid OA APC
were most often invoiced to authors (n=41,725; 58.2\%) and to a lesser
extent as part of agreements (n=24,250; 33.8\%). Interestingly, we also
found a small number of cases where hybrid APCs were waived (n=4,345;
6.1\%).

\begin{table}

\caption{\label{tab:invoicing_overview}Invoice Channels for Elsevier Hybrid OA Articles (2015-2019).}
\centering
\begin{tabular}[t]{lrr}
\toprule
Invoicing channel & Hybrid OA articles (n) & Percentage\\
\midrule
Author & 41,725 & 58.2\\
Agreement & 24,250 & 33.8\\
Fee waived & 4,345 & 6.1\\
Other & 1,323 & 1.8\\
\midrule
Total & 71,643 & 100.0\\
\bottomrule
\end{tabular}
\end{table}

\begin{figure}[H]
\caption{Elsevier Hybrid OA Articles by Invoice Channel and CC License (Per Year).}\label{fig:invoiceoverview}
{\centering \includegraphics[width=0.7\linewidth,]{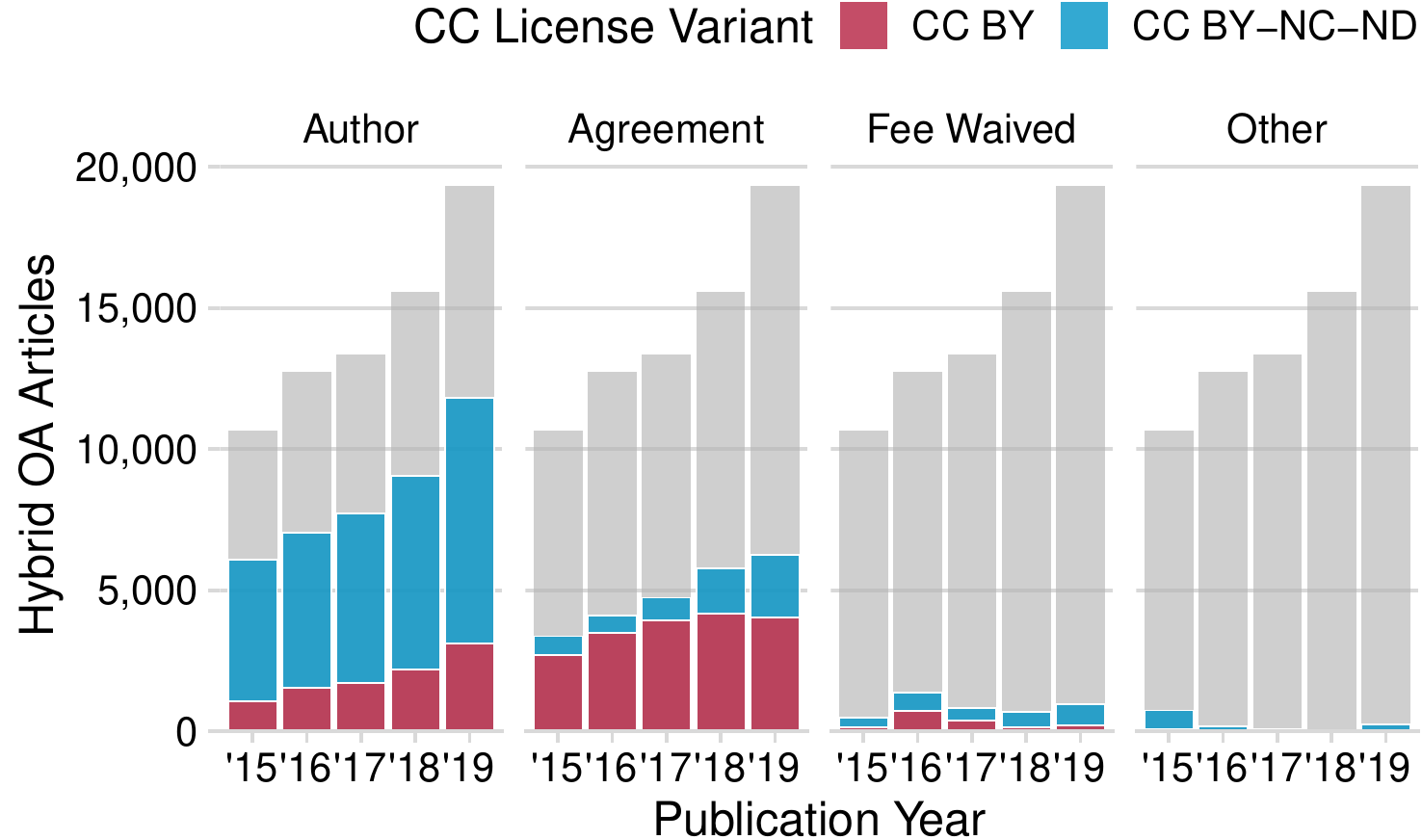} 

}

\end{figure}

Figure \ref{fig:invoiceoverview} illustrates that over the years,
Elsevier has increasingly invoiced authors directly
compared to research funders or academic consortia (``Agreement''). The
share of fee-waived articles remained relatively stable, but we found
different types of waivers. Around 51.7\% of fee waivers were linked to a third party. For
instance, the French Académie des Sciences,
presumably covering OA publication for 853 OA articles in its society journals for
affiliated authors. The remaining 48.3\% of waived articles did not
disclose any invoice recipient. Moreover, Figure \ref{fig:invoiceoverview}
compares the invoicing channels based on CC license variants.
When Elsevier invoiced authors directly, most OA articles in hybrid
journals were published under a non-commercial license (n=32,086;
76.9\%), whereas most articles billed as part of agreements were
licensed under the more permissive CC BY license (n=18,331; 75.6\%).

\vspace{1ex}

   {\raggedright \textit{Note:} Grey bars show the total number of hybrid OA articles published in Elsevier journals.\par}

We also observed large differences in OA invoicing among subject fields.
Table \ref{tab:invocing_by_subject} shows the number of OA articles by
subject field and invoice channel. For articles in nursing, decision
sciences, and pharmacology, toxicology and pharmaceutics, Elsevier
predominantly invoiced authors, whereas most energy and chemical
engineering articles were invoiced through agreements. Likewise, the
majority of articles in materials science, chemistry and physics and
astronomy were not invoiced to authors but facilitated through
agreements or waived. The large share of waived APCs in physics and
astronomy can be attributed to a single 2015 issue of Nuclear and
Particle Physics Proceedings.

\begin{table}[H]

\caption{\label{tab:invocing_by_subject}Invoice Channels by Subject (2015-2019).}
\centering
\resizebox{\linewidth}{!}{
\begin{tabular}[t]{lrrrrr}
\toprule
\multicolumn{2}{c}{ } & \multicolumn{4}{c}{Invoice channel (in \%)} \\
\cmidrule(l{3pt}r{3pt}){3-6}
Subject & OA articles & Author & Agreement & Fee waived & Other\\
\midrule
Nursing & 1,504 & \cellcolor[HTML]{DC3977}{82} & \cellcolor[HTML]{FFFFFF}{16} & \cellcolor[HTML]{FFF7E8}{1} & \cellcolor[HTML]{FCDE9C}{1}\\
Decision Sciences & 785 & \cellcolor[HTML]{E6576F}{72} & \cellcolor[HTML]{FCC088}{27} & \cellcolor[HTML]{FFFFFF}{0} & \cellcolor[HTML]{FCDE9C}{1}\\
Pharmacology, Toxicology and Pharmaceutics & 4,075 & \cellcolor[HTML]{EA646F}{69} & \cellcolor[HTML]{FCC088}{27} & \cellcolor[HTML]{FFF7E8}{1} & \cellcolor[HTML]{FAA476}{2}\\
Business, Management and Accounting & 1,751 & \cellcolor[HTML]{EC686E}{68} & \cellcolor[HTML]{FBAF7D}{29} & \cellcolor[HTML]{FFF0D2}{2} & \cellcolor[HTML]{FCDE9C}{1}\\
Dentistry & 288 & \cellcolor[HTML]{EC686E}{68} & \cellcolor[HTML]{FFFAF1}{17} & \cellcolor[HTML]{F89774}{10} & \cellcolor[HTML]{DC3977}{5}\\
Health Professions & 674 & \cellcolor[HTML]{ED6D6E}{67} & \cellcolor[HTML]{FCD898}{24} & \cellcolor[HTML]{FFE8BB}{3} & \cellcolor[HTML]{DC3977}{5}\\
Medicine & 23,623 & \cellcolor[HTML]{F0756E}{65} & \cellcolor[HTML]{FCD092}{25} & \cellcolor[HTML]{FCC98E}{6} & \cellcolor[HTML]{E34F6F}{4}\\
Agricultural and Biological Sciences & 7,956 & \cellcolor[HTML]{F38170}{63} & \cellcolor[HTML]{F9A075}{31} & \cellcolor[HTML]{FCD697}{5} & \cellcolor[HTML]{FCDE9C}{1}\\
Computer Science & 3,185 & \cellcolor[HTML]{F38170}{63} & \cellcolor[HTML]{F27F70}{36} & \cellcolor[HTML]{FFFFFF}{0} & \cellcolor[HTML]{FCDE9C}{1}\\
Economics, Econometrics and Finance & 2,071 & \cellcolor[HTML]{F48771}{62} & \cellcolor[HTML]{F79373}{33} & \cellcolor[HTML]{FCD697}{5} & \cellcolor[HTML]{FFFFFF}{0}\\
Veterinary & 2,091 & \cellcolor[HTML]{F58D72}{61} & \cellcolor[HTML]{FCD092}{25} & \cellcolor[HTML]{F1766E}{13} & \cellcolor[HTML]{FCDE9C}{1}\\
Social Sciences & 6,306 & \cellcolor[HTML]{F79273}{60} & \cellcolor[HTML]{F27F70}{36} & \cellcolor[HTML]{FFE8BB}{3} & \cellcolor[HTML]{FFFFFF}{0}\\
Earth and Planetary Sciences & 4,532 & \cellcolor[HTML]{F89874}{59} & \cellcolor[HTML]{F48671}{35} & \cellcolor[HTML]{FCD697}{5} & \cellcolor[HTML]{FFFFFF}{0}\\
Biochemistry, Genetics and Molecular Biology & 15,903 & \cellcolor[HTML]{F89874}{59} & \cellcolor[HTML]{F48671}{35} & \cellcolor[HTML]{FDE1A5}{4} & \cellcolor[HTML]{FAA476}{2}\\
Immunology and Microbiology & 5,542 & \cellcolor[HTML]{F89874}{59} & \cellcolor[HTML]{F58D72}{34} & \cellcolor[HTML]{FCC98E}{6} & \cellcolor[HTML]{FAA476}{2}\\
Environmental Science & 9,000 & \cellcolor[HTML]{F99D75}{58} & \cellcolor[HTML]{EF726E}{38} & \cellcolor[HTML]{FFE8BB}{3} & \cellcolor[HTML]{FCDE9C}{1}\\
Psychology & 3,087 & \cellcolor[HTML]{FAA376}{57} & \cellcolor[HTML]{EC686E}{40} & \cellcolor[HTML]{FFF0D2}{2} & \cellcolor[HTML]{FCDE9C}{1}\\
Neuroscience & 5,906 & \cellcolor[HTML]{FBAA7A}{56} & \cellcolor[HTML]{EC686E}{40} & \cellcolor[HTML]{FFE8BB}{3} & \cellcolor[HTML]{FCDE9C}{1}\\
Arts and Humanities & 1,565 & \cellcolor[HTML]{FCB882}{54} & \cellcolor[HTML]{E4536F}{44} & \cellcolor[HTML]{FFF7E8}{1} & \cellcolor[HTML]{FFFFFF}{0}\\
Mathematics & 2,105 & \cellcolor[HTML]{FCBF87}{53} & \cellcolor[HTML]{EF726E}{38} & \cellcolor[HTML]{FBAF7D}{8} & \cellcolor[HTML]{FFFFFF}{0}\\
Engineering & 8,405 & \cellcolor[HTML]{FCD395}{50} & \cellcolor[HTML]{E24C70}{46} & \cellcolor[HTML]{FFE8BB}{3} & \cellcolor[HTML]{FCDE9C}{1}\\
Materials Science & 4,838 & \cellcolor[HTML]{FCDA99}{49} & \cellcolor[HTML]{E14971}{47} & \cellcolor[HTML]{FDE1A5}{4} & \cellcolor[HTML]{FFFFFF}{0}\\
Chemistry & 4,033 & \cellcolor[HTML]{FCE0A1}{48} & \cellcolor[HTML]{E24C70}{46} & \cellcolor[HTML]{FCD697}{5} & \cellcolor[HTML]{FCDE9C}{1}\\
Energy & 4,286 & \cellcolor[HTML]{FEE7B8}{46} & \cellcolor[HTML]{DC3977}{52} & \cellcolor[HTML]{FFF7E8}{1} & \cellcolor[HTML]{FCDE9C}{1}\\
Chemical Engineering & 2,977 & \cellcolor[HTML]{FFEBC4}{45} & \cellcolor[HTML]{E14971}{47} & \cellcolor[HTML]{FCBC85}{7} & \cellcolor[HTML]{FCDE9C}{1}\\
Physics and Astronomy & 5,859 & \cellcolor[HTML]{FFFFFF}{40} & \cellcolor[HTML]{EF726E}{38} & \cellcolor[HTML]{DC3977}{22} & \cellcolor[HTML]{FFFFFF}{0}\\
\bottomrule
\end{tabular}}
\end{table}

\hypertarget{invoice-recipients}{%
\subsubsection*{Invoice Recipients}\label{invoice-recipients}}
\addcontentsline{toc}{subsubsection}{Invoice Recipients}

\begin{figure}[H]

\caption{Centrally Invoiced Elsevier Hybrid OA Articles by Country and CC License (2015-2019).}\label{fig:invoice_sponsor_country}

{\centering \includegraphics[width=0.7\linewidth,]{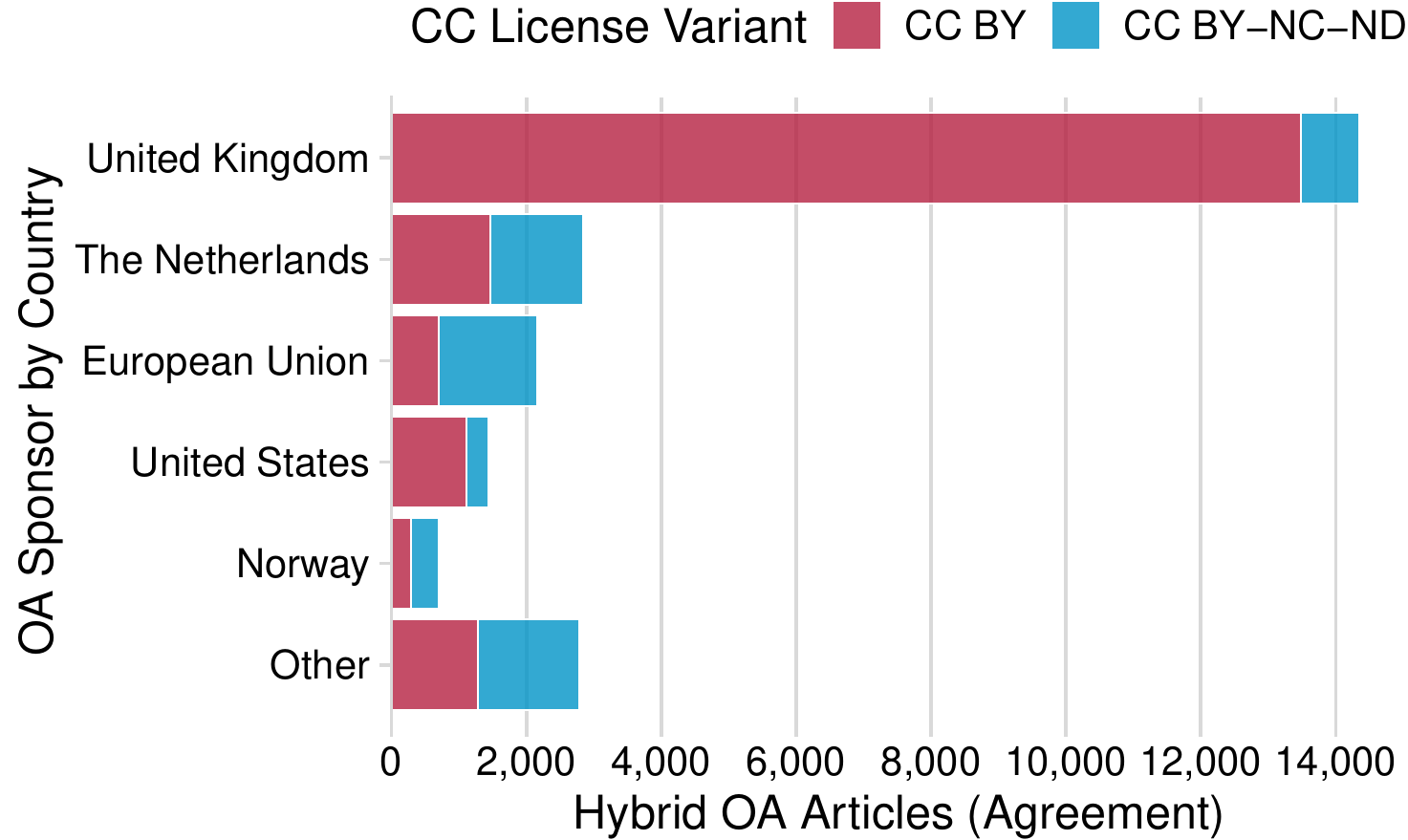} 

}

\end{figure}

Elsevier's data offer more insight into OA invoicing. Overall, we
identified 63 academic institutions and funders that received
invoices as part of publishing agreements. By a large margin, most invoices were issued to
UK-based research funders and institutions (n=14,344; 59.2\%), followed
by the Netherlands (n=2,835; 11.7\%) and the European Union (n=2,164;
8.9\%; EU). Figure \ref{fig:invoice_sponsor_country} presents the geographical distribution by license prevalence,
highlighting the dominance of CC BY licensed articles invoiced to institutions in the UK, the United States (US), and, to a much
lesser extent, the Netherlands. In contrast, most articles
invoiced to institutions from Norway or representing the EU
were published under a non-commercial license.

\begin{figure}[H]

\caption{Centrally Invoiced Elsevier Hybrid OA Articles by Country and Sector (Per Year).}\label{fig:sponsor_sector}

{\centering \includegraphics[width=0.7\linewidth,]{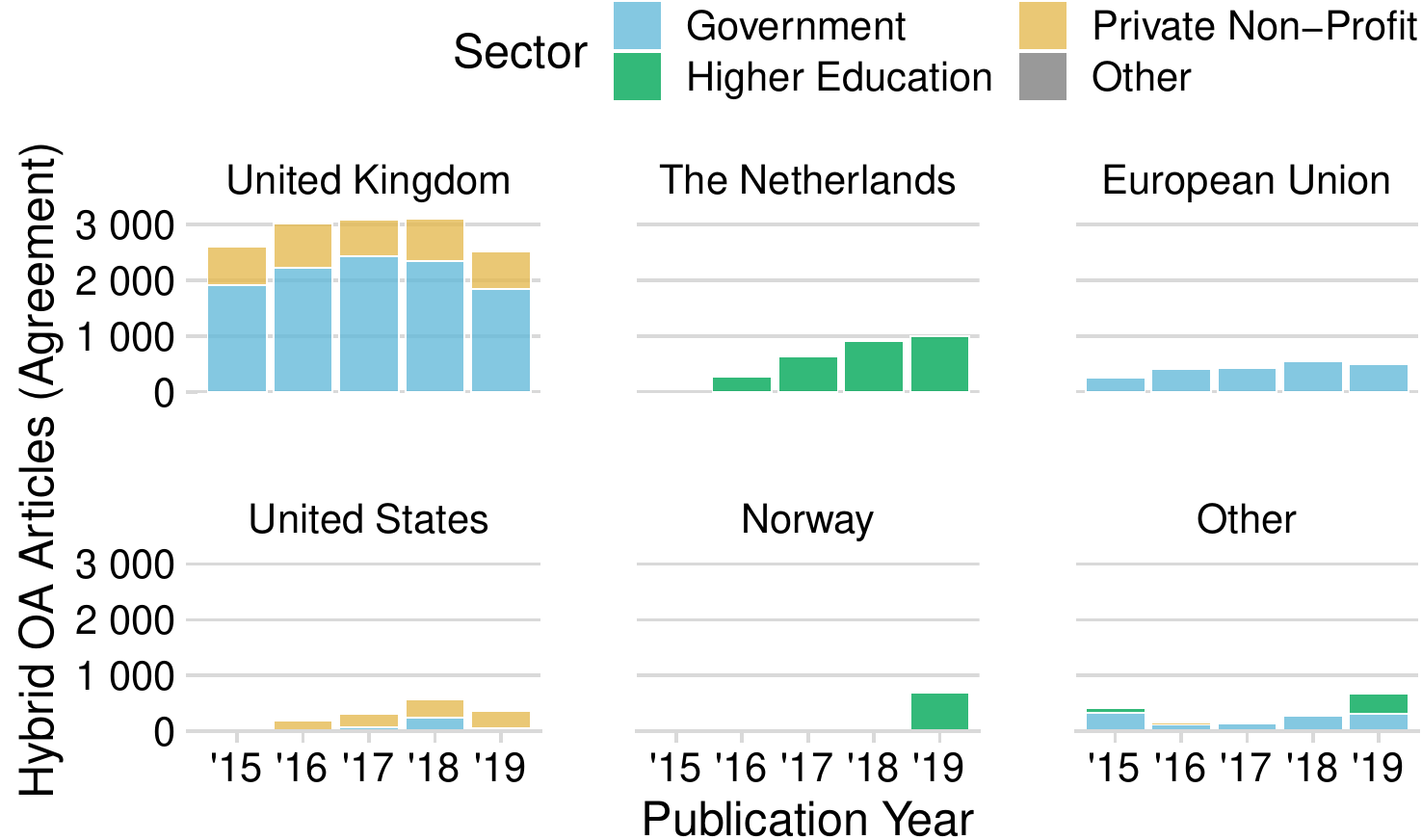} 

}

\end{figure}

Figure \ref{fig:sponsor_sector} shows the yearly distribution of centrally
invoiced articles by year and institutional sector. While UK-based OA
invoices were mainly addressed to discipline-specific governmental and
non-profit research funders, invoices to the Netherlands and Norway were
issued to national academic consortia representing the higher education
sector. In 2019, Elsevier also launched similar agreements in countries
with lower publication output including Hungary and Poland. Besides, we
found that invoice recipients from the UK and the US mainly
represented discipline-specific funders, while invoice recipients from other countries focused on a broad variety of
disciplines (see Table \ref{tab:subject_profile}).

\begin{table}

\caption{\label{tab:subject_profile}Share of Centrally Invoiced Elsevier Hybrid OA Articles by Country and Subject (2015-2019). }
\centering
\resizebox{\linewidth}{!}{
\begin{tabular}[t]{l|rrrrr|rl|rrrrr|rl|rrrrr|rl|rrrrr|rl|rrrrr|rl|rrrrr|rl|rrrrr|r}
\toprule
Country/Subject area & Health science & Life sciences & Physical sciences & SSH & Broad & Total\\
\midrule
United Kingdom & 25\% & 11\% & 21\% & 5\% & 0\% & 62\%\\
The Netherlands & 0\% & 0\% & 0\% & 0\% & 12\% & 12\%\\
European Union & 0\% & 0\% & 0\% & 0\% & 9\% & 9\%\\
United States & 6\% & 0\% & 0\% & 0\% & 1\% & 7\%\\
Norway & 0\% & 0\% & 0\% & 0\% & 3\% & 3\%\\
\addlinespace
Other & 1\% & 0\% & 1\% & 0\% & 5\% & 7\%\\
\midrule
Total & 32\% & 11\% & 22\% & 5\% & 30\% & 100\%\\
\midrule
\bottomrule
\end{tabular}}
\end{table}

\hypertarget{institutional-ranking}{%
\subsubsection*{Institutional ranking}\label{institutional-ranking}}
\addcontentsline{toc}{subsubsection}{Institutional ranking}

Table \ref{tab:sponsor_top_10} presents the top ten of 63 invoice recipients in our sample. Together, they accounted for around 80\% of
centrally invoiced APCs. The table also highlights the number of
distinct journals and the share of CC BY licensed articles. Notably,
UK-based research funders and the US-based Bill and Melinda Gates
Foundation mainly received invoices for CC BY licensed articles. In
contrast, articles invoiced to VSNU, the European Research Council, and
Norway Institutes had a much lower proportion of CC BY.

\begin{table}

\caption{\label{tab:sponsor_top_10}Elsevier Hybrid OA Invoice Recipients  (2015-2019).}
\centering
\resizebox{\linewidth}{!}{
\begin{tabular}[t]{lrrrrr}
\toprule
\multicolumn{4}{c}{ } & \multicolumn{2}{c}{Compliance (in \%)} \\
\cmidrule(l{3pt}r{3pt}){5-6}
Hybrid OA invoice recipients & Journals & Articles & \% & CC BY & OAPC\\
\midrule
Engineering and Physical Sciences Research Council & 619 & 4663 & 19 & \cellcolor[HTML]{DE3E75}{97} & \cellcolor[HTML]{E34E6F}{55}\\
VSNU & 357 & 2835 & 12 & \cellcolor[HTML]{FCC38A}{52} & \cellcolor[HTML]{FFFFFF}{0}\\
Wellcome Trust & 448 & 2506 & 10 & \cellcolor[HTML]{DD3D76}{98} & \cellcolor[HTML]{DD3D76}{66}\\
European Research Council & 541 & 1986 & 8 & \cellcolor[HTML]{FFFFFF}{32} & \cellcolor[HTML]{FFE9BD}{9}\\
Medical Research Council & 384 & 1922 & 8 & \cellcolor[HTML]{DF4274}{95} & \cellcolor[HTML]{E6596F}{51}\\
Natural Environment Research Council & 203 & 1357 & 6 & \cellcolor[HTML]{DE3E75}{97} & \cellcolor[HTML]{EC696E}{45}\\
Biotechnology and Biological Sciences Research Council & 309 & 1169 & 5 & \cellcolor[HTML]{E04573}{93} & \cellcolor[HTML]{F0736E}{41}\\
Bill and Melinda Gates Foundation & 254 & 1030 & 4 & \cellcolor[HTML]{DC3977}{100} & \cellcolor[HTML]{DC3977}{68}\\
Economic and Social Research Council & 245 & 922 & 4 & \cellcolor[HTML]{DE4075}{96} & \cellcolor[HTML]{E9616F}{48}\\
Norway Institutes & 374 & 694 & 3 & \cellcolor[HTML]{FFE9BD}{41} & \cellcolor[HTML]{FFFFFF}{0}\\
Other & 1012 & 5166 & 21 & \cellcolor[HTML]{FCBA84}{54} & \cellcolor[HTML]{FCBF87}{21}\\
\midrule
All & 1423 & 24250 & 100 & \cellcolor[HTML]{ED6C6E}{76} & \cellcolor[HTML]{F48671}{36}\\
\midrule
\bottomrule
\end{tabular}}
\end{table}

The table also highlights the proportion of APCs publicly
disclosed through the Open APC initiative, showing higher disclosure
rates for invoice recipients with a large CC BY share. 

\hypertarget{discussion}{%
\section{Discussion}\label{discussion}}

From 2015-2019, Elsevier recorded growth in the uptake of
hybrid OA: The number of hybrid OA
articles published per year doubled, the number of hybrid journals with
at least one OA article grew by 21\%, and the share of hybrid OA
articles relative to closed-accessed articles in these journals
increased from 2.6\% to 3.7\%. As Laakso \& Björk (2016), we found only
a weak relationship between the journals' SNIP and hybrid OA
uptake and observed disciplinary differences. In particular, we found the highest count of hybrid OA
articles in physical sciences journals (see Table
\ref{tab:subject_area_table}). This was followed by the life sciences
and health sciences, whereas the social sciences had the lowest count. This order mostly reflects the disciplines' overall
publication output. According to the Open Science Monitor (European Commission, 2019), the physical sciences publish the
most articles, followed by the health sciences, life sciences, and
social sciences.

Disciplinary differences in hybrid OA prevalence become more
meaningful when considering the relative share of OA articles to
closed-access articles. In line with previous research, we found that
Elsevier journals from the life sciences and social sciences (Jubb et
al., 2017; Kramer \& Bosman, 2018; Laakso \& Björk, 2016) recorded
greater than typical hybrid OA uptake, whereas physical sciences
journals generally had a lower than typical uptake (Kramer \& Bosman,
2018; Laakso \& Björk, 2016; Martín-Martín et al., 2018). In a
systematic review of disciplinary OA publishing patterns, Severin et al. (2020)
highlighted the importance of socio-cultural and technological
factors in shaping publishing cultures and practices.  For instance, since many
branches of the physical sciences have established self-archiving
practices (Severin et al. (2020), Björk et al. (2010), Laakso \& Björk
(2016)), researchers can provide OA through repositories and might perceive
much less of a need for hybrid OA. In contrast, hybrid OA in the life sciences could be enabled through project-based funding structures as they allow for easy integration of publishing costs (Severin et al. (2020).
On the other hand, the high hybrid OA uptake that we observed among the social sciences could point to the influential role of OA policies and invoicing agreements (Huang et al., 2020; Larivière \& Sugimoto, 2018).

Our invoicing data analysis found that most APCs were
invoiced to the author (n=41,725; 58.2\%). However, it is important to
emphasize although Elsevier's metadata classifies the sponsor type as
``author'', this does not necessarily mean that authors paid for APCs themselves but rather that APCs were not invoiced
through publishing agreements. It is possible that these APCs were covered through institutional funds or  research
grants. This would align with a recent Springer Nature survey that found hybrid
OA was predominantly supported through institutional and funder sources
(71\%), followed by OA agreements (34\%), while only 6\% were paid from
personal funds or savings (Monaghan et al., 2020). 

Further, we observed
notable differences in licensing. Most
hybrid OA articles invoiced to author were licensed under
the more restrictive CC BY-NC-ND license. Previous research, while lacking dedicated
studies on license selection,  suggests that authors tend to select more restrictive license variants when
given a choice (Fraser et al., 2020; Noorden, 2013; Rowley et
al., 2017). On the other hand, we found when hybrid OA was invoiced
through agreements, most articles were licensed under the more liberal
CC BY license. As several funding bodies mandate CC BY licenses,
including the UK research funders that account for 62\% of our
``agreement'' subsample, this result is perhaps not surprising but
suggests the effectiveness of such agreements.

Based on our findings, around a third of the articles were invoiced
 through OA agreements (e.g., research funders, national library
consortia). The predominance of UK funding bodies in this subsample reiterates reports from previous
studies that the UK's OA profile differs markedly from other countries
(cf.~Jubb et al. 2015; Jubb et al., 2017), pointing to the impact of science policy. To promote 
OA, the UK implemented centralized APC funding
and embraced hybrid OA as a transition model. Within five years of the
publication of the Finch Report in 2012, the UK recorded an 18\%
increase in immediate OA, coupled with a rise in hybrid OA from 2.7\% in
2012 to 15.4\% of all articles in 2016 (Jubb et al., 2017), which can be
attributed to the availability of funding from 
RCUK. Indeed, RCUK block grants have been the largest single source of
APC funds in the UK---the Wellcome Trust, and more recently
transformative agreements (Jubb et al., 2017; Tickel, 2018). However,
since then publishing expenditures of UK universities have been rising
rapidly (Jubb et al., 2017), so that several universities stopped supporting hybrid OA
in late 2018-2019 (University of Birmingham, n.d.; Walker, 2019). This development might
explain the slight decrease in hybrid articles invoiced to the UK our
data showed around that time (see Figure
\ref{fig:invoice_sponsor_country}).

Through this study we
demonstrated the utility and benefits of publisher-provided
metadata about hybrid OA invoicing and highlighted the need for extending
these to include comprehensive information about licensing and APC waivers. Metadata guidance should also consider the substantial amount of delayed OA content, which needs to be distinguished from hybrid OA.
Hence, our study substantiates the recommendations from the Efficiency and Standards for Article Charges (ESAC) Initiative that seek to increase efficiency and
transparency through improved invoicing and
reporting processes and metadata about OA funding (Geschuhn \& Stone, 2017).

While this study advances our knowledge about hybrid OA uptake and
invoicing, the limitations leave room for future research. We
focused on only one publisher, which limits the generalizability of our
findings because Elsevier's journal
portfolio, mix of business models, pricing, and promotion of various
options cannot be assumed to be representative of scholarly journal
publishing in general. Further, although Elsevier's invoicing data improves transparency, it seems likely that not all actual OA funding bodies are disclosed. For instance, most articles were invoiced to authors, but it remains unclear
if the APCs were paid by the authors themselves or
through institutional OA funds or research grants. Research into
this topic would improve our understanding of OA funding 
outside of publishing agreements. Moreover, our study demonstrates that
comprehensive mapping of the financial
flows of OA publishing requires complex, in-depth country-specific
analyses. Such studies would ideally draw on
various data sources to consider  research funders, the 
consortia landscape, OA policies, and  publishing agreements.

This study provides a snapshot of hybrid OA for Elsevier, the largest journal
publisher, prior to the impact of the implementation of Plan S, an initiative to accelerate the transition to OA that will no longer support hybrid OA (cOAlition S, n.d.). While many Plan S signatories have already had strong OA
policies, this harmonized approach is likely to affect publishing
decisions of funded authors, the licencing of their articles, and the
offerings and pricing of publishers at a larger scale than before. Because the new requirements apply to research funded from 2021 onward, a comparative study
on articles invoiced to Plan S signatories would be a fruitful
endeavor.

\hypertarget{conclusion}{%
\section{Conclusion}\label{conclusion}}

The primary aim of this empirical study was to investigate Elsevier's
hybrid OA publishing from 2015-2019 to better understand the volume and
invoicing of hybrid OA and to present a novel, data-driven approach for
such analyses. Our results indicate that although the number of hybrid
OA articles has increased over time, its uptake has remained low.
Notably, hybrid APCs were most often invoiced directly to the authors, followed by agreements, where only a few funding bodies were
the primary drivers of hybrid OA. Finally, our findings highlight that
publisher-provided metadata about the invoicing channels of (hybrid) OA
can facilitate research into and increase the understanding of the
financial flows of OA publishing.

Since the beginning, hybrid OA has been a challenging subject to
study due to the lack of standardized ways publishers flag such
content and APC funding data being limited self-reported data, surveys,
and other secondary sources. This study presented a novel
approach to studying APC invoicing that is based on publicly
available publisher-provided metadata, which can be used on on its own
or in combination with other public data sources to gain more detailed
and comprehensive insights into hybrid OA uptake and invoicing. If more publishers reported OA invoicing on the article
level and in a machine-readable format, this would increase transparency
and improve monitoring of the scholarly journal landscape over
time. As hybrid OA has  become a central element of OA
policies of research funders and libraries, consumer organizations could
require that invoicing information is added to the article-level
metadata. As long as publishers do not provide this data in a structured
and comprehensive format, they prevent benchmarking prices and therefore hinder competition. 

From recent science policy developments in Europe it appears that Big
Deals have gained support and remain firmly in place in the form of
transformative agreements. Through this study we can affirm that hybrid OA is complex
as the financial flows involve research funders, libraries, consortia, and authors. However, it is on
 publishers to increase the transparency of OA publishing, including hybrid OA and transformative agreements, by providing instantaneous open data about OA uptake and invoicing.

\hypertarget{data-availability}{%
\section{Data Availability}\label{data-availability}}

The source code data analysis is available on GitHub:
\url{https://github.com/njahn82/elsevier_hybrid_invoicing}

\hypertarget{references}{%
\section*{References}\label{references}}
\addcontentsline{toc}{section}{References}

\hypertarget{refs}{}
\begin{cslreferences}

\leavevmode\hypertarget{ref-Bergstrom_2014}{}%
Bergstrom, T. C., Courant, P. N., McAfee, R. P., \& Williams, M. A.
(2014). Evaluating Big Deal journal bundles. \emph{Proceedings of the
National Academy of Sciences}, \emph{111}(26), 9425--9430.
\url{https://doi.org/10.1073/pnas.1403006111}

\leavevmode\hypertarget{ref-Bj_rk_2012}{}%
Björk, B.-C. (2012). The hybrid model for open access publication of
scholarly articles: A failed experiment? \emph{Journal of the American
Society for Information Science and Technology}, \emph{63}(8),
1496--1504. \url{https://doi.org/10.1002/asi.22709}

\leavevmode\hypertarget{ref-Bj_rk_2010}{}%
Björk, B.-C., Welling, P., Laakso, M., Majlender, P., Hedlund, T., \&
Guðnason, G. (2010). Open access to the scientific journal literature:
Situation 2009. \emph{PLoS ONE}, \emph{5}(6), e11273.
\url{https://doi.org/10.1371/journal.pone.0011273}

\leavevmode\hypertarget{ref-Borrego_2020}{}%
Borrego, Á., Anglada, L., \& Abadal, E. (2020). Transformative
agreements: Do they pave the way to open access? \emph{Learned
Publishing}. \url{https://doi.org/10.1002/leap.1347}

%\leavevmode\hypertarget{ref-Castro_2019}{}%
%Castro, P. de. (2019). \emph{Running a no-hybrid open access funding
%policy: Some results}. Open Access Strathclyde.
%\url{https://strathoa.tumblr.com/post/185703709380/running-a-no-hybrid-open-access-funding-policy}

\leavevmode\hypertarget{ref-crminer}{}%
Chamberlain, S. (2020). \emph{crminer: Fetch scholary full text from
CrossRef}. \url{https://CRAN.R-project.org/package=crminer}

\leavevmode\hypertarget{ref-rcrossref}{}%
Chamberlain, S., Zhu, H., Jahn, N., Boettiger, C., \& Ram, K. (2020).
\emph{rcrossref: Client for various CrossRef APIs}.
\url{https://CRAN.R-project.org/package=rcrossref}

%\leavevmode\hypertarget{ref-Chawla_2020}{}%
%Chawla, D. (2020). This tool is saving universities millions of dollars
%in journal subscriptions. \emph{Science}.
%\url{https://doi.org/10.1126/science.abd7483}

\leavevmode\hypertarget{ref-Plan_s}{}%
cOAlition S (n.d.). \emph{Plan S: Principles and implementation}.
\url{https://web.archive.org/web/20210129165306/https://www.coalition-s.org/addendum-to-the-coalition-s-guidance-on-the-implementation-of-plan-s/principles-and-implementation/}.

\leavevmode\hypertarget{ref-Crossref_2020}{}%
Crossref. (2020). \emph{March 2020 public data file from crossref}.
Crossref. \url{https://doi.org/10.13003/83b2gp}

\leavevmode\hypertarget{ref-Els_Agreements}{}%
Elsevier. (n.d.-a). \emph{Agreements}.
\url{https://web.archive.org/web/20210127152729/https://www.elsevier.com/open-access/agreements}.

\leavevmode\hypertarget{ref-Els_Archive}{}%
Elsevier. (n.d.-b). \emph{Open archive}.
\url{http://web.archive.org/web/20210127201740/https://www.elsevier.com/open-access/open-archive}.

%\leavevmode\hypertarget{ref-Els_videos}{}%
%Elsevier. (n.d.-c). \emph{Publishing Journey videos: How do I complete
%the Rights and Access form?}
%\url{https://web.archive.org/web/20210127154144/https://service.elsevier.com/app/answers/detail/a_id/29789/supporthub/publishing/track/APN2ZgoIDv8a~RNiGvwa~yKgpv0qOS75Mv9e~zj~PP_X/}.

%\leavevmode\hypertarget{ref-lund}{}%
%Elsevier. (n.d.-d). \emph{Publishing options: BIBSAM institute
%associated authors}.
%\url{https://web.archive.org/web/20210127153520/https://www.ub.lu.se/en/sites/ub.lu.se.en/files/publicering_under_elsevier_avtalet.pdf}.

\leavevmode\hypertarget{ref-pricing}{}%
Elsevier. (n.d.-c). \emph{Pricing}.
\url{https://web.archive.org/web/20210127152857/https://www.elsevier.com/about/policies/pricing}.

\leavevmode\hypertarget{ref-vsnu}{}%
Elsevier. (n.d.-d). \emph{Publishing options for: Dutch universities \&
institutes associated authors}.
\url{https://web.archive.org/web/20190530214833/https://www.openaccess.nl/sites/www.openaccess.nl/files/documenten/elsevier_-_vsnu_workflow_updated_18july_v2.pdf}.

\leavevmode\hypertarget{ref-OS_Monitor}{}%
European Commission. (2019). \emph{Open science monitor: Trends for open
access to publications}.
\url{https://ec.europa.eu/info/research-and-innovation/strategy/goals-research-and-innovation-policy/open-science/open-science-monitor/trends-open-access-publications_en}

\leavevmode\hypertarget{ref-Fraser_2020}{}%
Fraser, N., Brierley, L., Dey, G., Polka, J. K., Pálfy, M., Nanni, F.,
\& Coates, J. A. (2020). \emph{Preprinting the COVID-19 pandemic}.
bioRxiv. \url{https://doi.org/10.1101/2020.05.22.111294}

\leavevmode\hypertarget{ref-Frazier_2001}{}%
Frazier, K. (2001). The librarians' dilemma: Contemplating the costs of
the "Big Deal". \emph{D-Lib Magazine}, \emph{7}(3).
\url{https://web.archive.org/web/20210125073550/https://librarytechnology.org/document/8950}

\leavevmode\hypertarget{ref-Geschuhn_2017}{}%
Geschuhn, K., \& Stone, G. (2017). It's the workflows, stupid! What is
required to make ``offsetting'' work for the open access transition.
\emph{Insights the UKSG Journal}, \emph{30}(3), 103--114.
\url{https://doi.org/10.1629/uksg.391}

\leavevmode\hypertarget{ref-Graaf_2017}{}%
Graaf, M. V. D. (2017). \emph{Paying for open access: The author's
perspective}. Zenodo. \url{https://doi.org/10.5281/ZENODO.438037}

\leavevmode\hypertarget{ref-Harrison_2019}{}%
Harrison, P. (2019). \emph{What are mirror journals, and can they offer
a new world of open access?} Elsevier B.V.
\url{https://web.archive.org/web/20210109033031/https://www.elsevier.com/connect/what-are-mirror-journals-and-can-they-offer-a-new-world-of-open-access}

\leavevmode\hypertarget{ref-Hendricks_2020}{}%
Hendricks, G., Tkaczyk, D., Lin, J., \& Feeney, P. (2020). Crossref: The
sustainable source of community-owned scholarly metadata.
\emph{Quantitative Science Studies}, \emph{1}(1), 414--427.
\url{https://doi.org/10.1162/qss_a_00022}

\leavevmode\hypertarget{ref-Hinchliffe_2019}{}%
Hinchliffe, L. J. (2019). \emph{Transformative agreements: A primer}.
\url{https://web.archive.org/web/20210128170342/https://scholarlykitchen.sspnet.org/2019/04/23/transformative-agreements/};
The Scholarly Kitchen.

\leavevmode\hypertarget{ref-Huang_2020}{}%
Huang, C.-K. (Karl), Neylon, C., Hosking, R., Montgomery, L., Wilson, K.
S., Ozaygen, A., \& Brookes-Kenworthy, C. (2020). Evaluating the impact
of open access policies on research institutions. \emph{eLife},
\emph{9}. \url{https://doi.org/10.7554/elife.57067}

\leavevmode\hypertarget{ref-Jahn_2016}{}%
Jahn, N., \& Tullney, M. (2016). A study of institutional spending on
open access publication fees in germany. \emph{PeerJ}, \emph{4}, e2323.
\url{https://doi.org/10.7717/peerj.2323}

\leavevmode\hypertarget{ref-Jubb_2015}{}%
Jubb, M., Goldstein, S., Amin, M., Plume, A., Oeben, S., Aisati, M.,
Pinfield, S., Bath, P., Salter, J., Johnson, R., \& Fosci, M. (2015).
\emph{Monitoring the transition to open access: A report for the
universities uk open access co-ordination group}.
\url{https://web.archive.org/web/20190523005240/https://www.acu.ac.uk/research-information-network/monitoring-transition-to-open-access}

\leavevmode\hypertarget{ref-Jubb_2017}{}%
Jubb, M., Plume, A., Oeben, S., Brammer, L., Johnson, R., Bütün, C., \&
Pinfield, S. (2017). \emph{Monitoring the transition to open access:
December 2017}.
\url{https://www.universitiesuk.ac.uk/policy-and-analysis/reports/Documents/2017/monitoring-transition-open-access-2017.pdf}

\leavevmode\hypertarget{ref-Kirkman_2018}{}%
Kirkman, N. S. (2018). \emph{A study of open access publishing by NHMRC 
grant recipients} {[}Curtin University{]}.
\url{http://hdl.handle.net/20.500.11937/77026}

\leavevmode\hypertarget{ref-Kramer_2018}{}%
Kramer, B., \& Bosman, J. (2018). \emph{Towards a Plan S gap analysis:
open access potential across disciplines using Web of Science and DOAJ}
{[}Data set{]}. Zenodo. \url{https://doi.org/10.5281/zenodo.1979937}

\leavevmode\hypertarget{ref-Laakso_2016}{}%
Laakso, M., \& Björk, B.-C. (2016). Hybrid open access--a longitudinal
study. \emph{Journal of Informetrics}, \emph{10}(4), 919--932.
\url{https://doi.org/10.1016/j.joi.2016.08.002}

\leavevmode\hypertarget{ref-Larivi_re_2015}{}%
Larivière, V., Haustein, S., \& Mongeon, P. (2015). The oligopoly of
academic publishers in the digital era. \emph{PLOS ONE}, \emph{10}(6),
e0127502. \url{https://doi.org/10.1371/journal.pone.0127502}

\leavevmode\hypertarget{ref-Larivi_re_2018}{}%
Larivière, V., \& Sugimoto, C. R. (2018). Do authors comply when funders
enforce open access to research? \emph{Nature}, \emph{562}(7728),
483--486. \url{https://doi.org/10.1038/d41586-018-07101-w}

%\leavevmode\hypertarget{ref-Lawson_2018}{}%
%Lawson, S. (2017). \emph{Report on offset agreements: Evaluating %current
%JISC collections deals. Year 2 -- evaluating 2016 deals}. figshare.
%\url{https://doi.org/10.6084/M9.FIGSHARE.5383861.V1}

\leavevmode\hypertarget{ref-Lawson_2015}{}%
Lawson, S. (2015). "Total cost of ownership" of scholarly communication:
Managing subscription and APC payments together. \emph{Learned
Publishing}, \emph{28}(1), 9--13. \url{https://doi.org/10.1087/20150103}

\leavevmode\hypertarget{ref-Lawson_2016}{}%
Lawson, S., Gray, J., \& Mauri, M. (2016). Opening the black box of
scholarly communication funding: A public data infrastructure for
financial flows in academic publishing. \emph{Open Library of
Humanities}, \emph{2}(1). \url{https://doi.org/10.16995/olh.72}

Marques, M., \& Stone, G. (2020). Transitioning to open access: An evaluation of the UK Springer Compact Agreement Pilot 2016–2018. \emph{College \& Research Libraries}, 81(6), 913–927. \url{https://doi.org/10.5860/crl.81.6.913}

\leavevmode\hypertarget{ref-Marques_2019}{}%
Marques, M., Woutersen-Windhouwer, S., \& Tuuliniemi, A. (2019).
Monitoring agreements with open access elements: Why article-level
metadata are important. \emph{Insights the UKSG Journal}, \emph{32}.
\url{https://doi.org/10.1629/uksg.489}

\leavevmode\hypertarget{ref-Mart_n_Mart_n_2018}{}%
Martín-Martín, A., Costas, R., Leeuwen, T. van, \& López-Cózar, E. D.
(2018). Evidence of open access of scientific publications in google
scholar: A large-scale analysis. \emph{Journal of Informetrics},
\emph{12}(3), 819--841. \url{https://doi.org/10.1016/j.joi.2018.06.012}

\leavevmode\hypertarget{ref-Matthias_2020}{}%
Matthias, L. (2020). \emph{Publisher OA portfolios 2.0} (Version 2.0)
{[}Data set{]}. Zenodo. \url{https://doi.org/10.5281/zenodo.3841568}

\leavevmode\hypertarget{ref-Matthias_2019}{}%
Matthias, L., Jahn, N., \& Laakso, M. (2019). The two-way street of open
access journal publishing: Flip it and reverse it. \emph{Publications},
\emph{7}(2), 23. \url{https://doi.org/10.3390/publications7020023}

\leavevmode\hypertarget{ref-Mittermaier_2015}{}%
Mittermaier, B. (2015). Double dipping in hybrid open access -- chimera
or reality? \emph{ScienceOpen Research}.
\url{https://doi.org/10.14293/s2199-1006.1.sor-socsci.aowntu.v1}

\leavevmode\hypertarget{ref-Monaghan_2020}{}%
Monaghan, J., Lucraft, M., \& Allin, K. (2020). \emph{'APCs in the
wild': Could increased monitoring and consolidation of funding
accelerate the transition to open access?} figshare.
\url{https://doi.org/10.6084/M9.FIGSHARE.11988123.V4}

\leavevmode\hypertarget{ref-Nelson_2017}{}%
Nelson, G. M., \& Eggett, D. L. (2017). Citations, mandates, and money:
Author motivations to publish in chemistry hybrid open access journals.
\emph{Journal of the Association for Information Science and
Technology}, \emph{68}(10), 2501--2510.
\url{https://doi.org/10.1002/asi.23897}

\leavevmode\hypertarget{ref-Van_Noorden_2013}{}%
Noorden, R. V. (2013). Researchers opt to limit uses of open-access
publications. \emph{Nature}.
\url{https://doi.org/10.1038/nature.2013.12384}

\leavevmode\hypertarget{ref-Frascati}{}%
OECD. (2015). \emph{Frascati manual 2015: Guidelines for collecting and
reporting data on research and experimental development}. OECD
Publishing. \url{https://doi.org/10.1787/24132764}

%\leavevmode\hypertarget{ref-Olsson1271866}{}%
%Olsson, L. (2018). \emph{Evaluation of offset agreements -- report 4 :
%Springer compact} (p. 20). Kungliga biblioteket.
%\url{http://urn.kb.se/resolve?urn=urn\%3Anbn\%3Ase\%3Ahb\%3Adiva-15501}

\leavevmode\hypertarget{ref-Pieper_2018}{}%
Pieper, D., \& Broschinski, C. (2018). OpenAPC: A contribution to a
transparent and reproducible monitoring of fee-based open access
publishing across institutions and nations. \emph{Insights the UKSG
Journal}, \emph{31}. \url{https://doi.org/10.1629/uksg.439}

\leavevmode\hypertarget{ref-Pinfield_2016}{}%
Pinfield, S., Salter, J., \& Bath, P. A. (2016). The "total cost of
publication" in a hybrid open-access environment: Institutional
approaches to funding journal article-processing charges in combination
with subscriptions. \emph{Journal of the Association for Information
Science and Technology}, \emph{67}(7), 1751--1766.
\url{https://doi.org/10.1002/asi.23446}

\leavevmode\hypertarget{ref-Piwowar_2018}{}%
Piwowar, H., Priem, J., Larivière, V., Alperin, J. P., Matthias, L.,
Norlander, B., Farley, A., West, J., \& Haustein, S. (2018). The state
of OA: A large-scale analysis of the prevalence and impact of open
access articles. \emph{PeerJ}, \emph{6}, e4375.
\url{https://doi.org/10.7717/peerj.4375}

\leavevmode\hypertarget{ref-Piwowar_2019}{}%
Piwowar, H., Priem, J., \& Orr, R. (2019). \emph{The future of OA: A
large-scale analysis projecting open access publication and readership}.
bioRxiv. \url{https://doi.org/10.1101/795310}

\leavevmode\hypertarget{ref-P_l_nen_2020}{}%
Pölönen, J., Laakso, M., Guns, R., Kulczycki, E., \& Sivertsen, G.
(2020). Open access at the national level: A comprehensive analysis of
publications by finnish researchers. \emph{Quantitative Science
Studies}, \emph{1}(4), 1396--1428.
\url{https://doi.org/10.1162/qss_a_00084}

\leavevmode\hypertarget{ref-Prosser_2015}{}%
Prosser, D. (2015). \emph{The costs of double dipping}.
\url{https://web.archive.org/web/20210128164930/https://www.rluk.ac.uk/the-costs-of-double-dipping/}.

\leavevmode\hypertarget{ref-Prosser_2003}{}%
Prosser, D. C. (2003). From here to there: A proposed mechanism for
transforming journals from closed to open access. \emph{Learned
Publishing}, \emph{16}(3), 163--166.
\url{https://doi.org/10.1087/095315103322110923}

\leavevmode\hypertarget{ref-r}{}%
R Core Team. (2020). \emph{R: A language and environment for statistical
computing}. R Foundation for Statistical Computing.
\url{https://www.R-project.org/}

\leavevmode\hypertarget{ref-Robinson_Garcia_2020}{}%
Robinson-Garcia, N., Costas, R., \& Leeuwen, T. N. van. (2020). Open
access uptake by universities worldwide. \emph{PeerJ}, \emph{8}, e9410.
\url{https://doi.org/10.7717/peerj.9410}

\leavevmode\hypertarget{ref-Rowley_2017}{}%
Rowley, J., Johnson, F., Sbaffi, L., Frass, W., \& Devine, E. (2017).
Academics' behaviors and attitudes towards open access publishing
in scholarly journals. \emph{Journal of the Association for Information
Science and Technology}, \emph{68}(5), 1201--1211.
\url{https://doi.org/10.1002/asi.23710}

\leavevmode\hypertarget{ref-Schimmer_2015}{}%
Schimmer, R., Geschuhn, K., \& Vogler, A. (2015). \emph{Disrupting the
subscription journals' business model for the necessary large-scale
transformation to open access}. Max Planck Digital Library.
\url{https://doi.org/10.17617/1.3}

\leavevmode\hypertarget{ref-Severin_2020}{}%
Severin, A., Egger, M., Eve, M. P., \& Hürlimann, D. (2020).
Discipline-specific open access publishing practices and barriers to
change: An evidence-based review. \emph{F1000Research}, \emph{7}.
\url{https://doi.org/10.12688/f1000research.17328.2}

\leavevmode\hypertarget{ref-Shieber_2009}{}%
Shieber, S. M. (2009). Equity for open-access journal publishing.
\emph{PLoS Biology}, \emph{7}(8), e1000165.
\url{https://doi.org/10.1371/journal.pbio.1000165}

\leavevmode\hypertarget{ref-Tickel_2018}{}%
Tickel, A. (2018). \emph{Open access to research publications -- 2018}.
\url{https://assets.publishing.service.gov.uk/government/uploads/system/uploads/attachment_data/file/774956/Open-access-to-research-publications-2018.pdf}

\leavevmode\hypertarget{ref-birmingham}{}%
University of Birmingham. (n.d.). \emph{UKRI open access block grant}.
\url{https://intranet.birmingham.ac.uk/as/libraryservices/library/research/open-access/funding/ukri-open-access-block-grant.aspx}

%\leavevmode\hypertarget{ref-oxford_2019}{}%
%University of Oxford. (2019). \emph{Change to oxford's policy for rcuk
%oa block grant 2nd december 2019}.
%\url{http://openaccess.ox.ac.uk/2019/12/02/change-to-oxfords-policy-for-rcuk-oa-block-grant-2nd-december-2019/}

\leavevmode\hypertarget{ref-Unpaywall_para}{}%
Unpaywall. (n.d.-a). \emph{What does is\_paratext mean in the API?}
\url{https://web.archive.org/web/20201126131640/https://support.unpaywall.org/support/solutions/articles/44001894783}.

\leavevmode\hypertarget{ref-Unpaywall_types}{}%
Unpaywall. (n.d.-b). \emph{What do the types of oa\_status (green, gold,
hybrid, and bronze) mean?}
\url{http://web.archive.org/web/20210127192910/https://support.unpaywall.org/support/solutions/articles/44001777288}.

\leavevmode\hypertarget{ref-Unpaywall_oa_license}{}%
Unpaywall. (n.d.-c). \emph{What is an OA license?}
\url{http://web.archive.org/web/20210127192625/https://support.unpaywall.org/support/solutions/articles/44002063718-what-is-an-oa-license-}.

\leavevmode\hypertarget{ref-walker_2019}{}%
Walker, D. (2019). \emph{Research Councils UK open access funding
2019-2020}. Library \& Archives Service at The London School of Hygiene
\& Tropical Medicine.
\url{https://blogs.lshtm.ac.uk/library/2019/03/05/research-councils-uk-open-access-funding-2019-2020/}

\leavevmode\hypertarget{ref-tidyverse}{}%
Wickham, H., Averick, M., Bryan, J., Chang, W., McGowan, L. D.,
François, R., Grolemund, G., Hayes, A., Henry, L., Hester, J., Kuhn, M.,
Pedersen, T. L., Miller, E., Bache, S. M., Müller, K., Ooms, J.,
Robinson, D., Seidel, D. P., Spinu, V., \ldots{} Yutani, H. (2019).
Welcome to the tidyverse. \emph{Journal of Open Source Software},
\emph{4}(43), 1686. \url{https://doi.org/10.21105/joss.01686}

\end{cslreferences}

\end{document}